\documentclass[a4paper,fleqn]{cas-dc}

\usepackage[numbers]{natbib}
\usepackage{doi}

\ExplSyntaxOn
\keys_set:nn { stm / mktitle } { nologo }
\ExplSyntaxOff

\begin{document}
\def\floatpagepagefraction{1}
\def\textpagefraction{.001}

\shorttitle{$\beta$-particle measurements for cross-section determination}    

\shortauthors{S. R. Kov\'acs, G. G. Kiss} 

\title[mode = title]{Measurement of \texorpdfstring{$\beta$}{beta}-particles to determine cross sections relevant to the weak r-process}



%
\author[1,2]{S\'andor R. Kov\'acs}[
       orcid=0009-0009-9796-2970,
       ]



\author[2]{Tibor Norbert Szegedi}[
	]

\author[1,2]{\'Akos T\'oth}[
       ]

\author[3]{Attila N\'emeth}[
       ]

\author[3]{Manoj Kumar Pal}[
       orcid=0000-0003-2803-7449,
       ]

\author[3]{Edit Szil\'agyi}[
       orcid=0000-0003-3210-0121,
       ]
			
\author[2]{Mih\'aly Braun}[
       ]
			
\author[2]{Gy\"orgy Gy\"urky}[
       orcid=0000-0003-3779-6463,
       ]

\author[2]{Zolt\'an Elekes}[
       orcid=0000-0003-0571-8719,
       ]
       
\author[2]{Zolt\'an Hal\'asz}[
       orcid=0000-0002-4678-5232,
       ]
\author[2]{Tam\'as Sz\"ucs}[
       orcid=0000-0002-9980-4847,
       ]		
							
\author[2]{G\'abor Gyula Kiss}[type=editor,
       orcid=0000-0002-6872-916X,
       ]
\cormark[2]
\ead{ggkiss@atomki.hu}

\affiliation[1]{organization={University of Debrecen, Faculty of Science and Technology, Doctoral School of Physics},
            addressline={Egyetem t\'er 1}, 
            city={Debrecen},
            citysep={}, 
            postcode={H-4002}, 
            country={Hungary}}

\affiliation[2]{organization={HUN-REN Institute for Nuclear Research (ATOMKI)},
            addressline={Bem t\'er 18/c}, 
            city={Debrecen},
            postcode={4026}, 
            country={Hungary}}

\affiliation[3]{organization={HUN-REN Wigner Research Centre for Physics},
            addressline={Konkoly-Thege Mikl\'os \'ut 29-33.}, 
            city={Budapest},
            postcode={1121}, 
            country={Hungary}}

\begin{abstract}
       The neutron-rich isotopes with 30 $\leq$ Z $\leq$ 45 are thought to be synthesised in neutrino-driven winds after the collapse of a massive star.             
       This nucleosynthesis scenario, called the weak r-process, is studied using nuclear reaction network calculations. 
       The accuracy of the nucleosynthesis simulations is strongly influenced by the reliability of the nuclear physics input parameters used. 
       Recently, it has been demonstrated that ($\alpha$,n) reactions play a particularly important role in the weak r-process, but their rates -- computed from the cross sections -- are only known with large uncertainties in the astrophysically relevant temperature range. 
       The half-lives of the products of some key reactions are such that, in principle, the cross sections can be studied using the activation technique. 
       In many cases, however, the $\beta$-decay of the reaction products leads to the ground state of the daughter nucleus, hence no gamma emission occurs. 
       The purpose of this manuscript is to present our setup with which we determine the cross sections by measuring the yield of $\beta$-particles emitted during radioactive decay.
       The $^{86}$Kr($\alpha$,n)$^{89}$Sr reaction cross-section measurement is used, as a case study. 
\end{abstract}


\maketitle



\section{Introduction}\label{weak}

       It has been known for more than 70 years that neutron capture processes play a key role in the synthesis of the chemical elements heavier than iron \cite{Bur57,Cam57}. 
       Depending on the neutron density of the astrophysical environment, two significantly different neutron-capture processes are distinguished, which are called slow (s-) and rapid (r-) processes \cite{Cow21,Lug23}. 
       In the first case, the neutron density is low (about 10$^{6-10}$ neutron/cm$^3$), the lifetime of the nuclei synthesised by neutron capture is comparable or shorter than the time elapsed between successive neutron captures, and consequently the process moves in the valley of stability. 
       On the contrary, in the r-process, due to the extremely high neutron density (exceeding $\rho_n$ $\approx$ 10$^{26}$ neutron/cm$^3$), the path of nucleosynthesis moves far away from the valley of stability through series of neutron captures, and the stable isotopes are built up by consecutive $\beta$-decays when the neutron flux ceases.
       However, it is now also known that in order to describe the observed abundance of the isotopes heavier than iron, it is necessary to assume additional nucleosynthesis processes.

       Indeed, recently it has been shown that lower mass neutron-rich isotopes can be synthesised in neutrino-driven winds following the collapse of a massive star \cite{Qia07,Mey92,Arc11a,Arc14}. 
       The modelling of this so-called weak r-process nucleosynthesis scenario requires the use of an extended network of several thousand reactions. 
       The required reaction rates are calculated from the cross sections computed with the Hauser-Feshbach statistical model, which relies on nuclear physics inputs. 
       Recently, a series of sensitivity calculations have been performed to evaluate the theoretical uncertainty of these cross-section calculations \cite{Per16,Moh16,Bli17}. 
       These works have shown that the main source of uncertainty is the use of different $\alpha$-nucleus optical potential parameter sets ($\alpha$-OMP's).
       With decreasing interaction energy, the difference between the cross-section predictions calculated with different $\alpha$-OMP's increases drastically, often exceeding an order of magnitude in the astrophysically relevant energy region. 
       It is therefore crucial to determine the cross sections at low energies, preferably in- or as close to the astrophysically relevant energy range as possible.

       The well-known activation technique \cite{Gyu19} proved to be successful for such cross-section measurements. 
       Recently, the $^{96}$Zr($\alpha$,n)$^{99}$Mo and the $^{100}$Mo($\alpha$,n)$^{103}$Ru reactions have been measured at energies partially covering the astrophysically relevant region \cite{Kis21,Sze21,Ham22}. 
       However, the applicability of the activation technique is limited. 
       This technique can be mainly used to study reactions where the half-life of the product is adequate (typically between a few tens of minutes and a few weeks) and its decay is accompanied by a radiation that can be easily measured.
       Generally, as in the case of the above-mentioned reactions, the $\gamma$-radiation emitted after $\beta$-decay is measured with HPGe detectors. 
       However, $\beta$-decay sometimes leads directly to the ground state of the daughter nucleus and thus no $\gamma$-radiation is emitted. 
       Such reactions, with important role in the weak r-process network, are for example the $^{86}$Kr($\alpha$,n)$^{89}$Sr and $^{87}$Rb($\alpha$,n)$^{90}$Y \footnotemark \cite{Psa22}. 
       For the cross section measurement of these reactions, the number of reaction products can be determined, for example, by measuring the emitted $\beta$-particles. 
       In this manuscript, we present the setup built for this purpose at the HUN-REN Institute for Nuclear Research (ATOMKI) and the results of the calibration and validation measurements. 

       \footnotetext{Note that in both reactions, the $\beta$-decay is followed by low relative intensity gamma radiation, which does not provide high enough yield for a cross section measurement}

\section {Target preparation and characterization}\label{sec:target}

       An activation experiment can generally be divided into two separate phases: the irradiation phase, in which the reaction products are produced, and the measurement of the induced activity \cite{Gyu19}. 
       Since krypton is a noble gas, targets prepared by ion-implantation or gas targets can be used for the irradiation. 
       In order to limit the possible systematic uncertainties, both types of targets have been used in the present work and their properties are discussed in details in the next sections. 
       Our experimental technique is presented using the $^{86}$Kr($\alpha$,n)$^{89}$Sr reaction as an example.
       The reaction Q-value is -2.7 MeV and the region in which precise data is needed for astrophysical calculation lies between 5-9 MeV depending on the temperature of the environment.
       The half life of $^{89}$Sr is T$_{1/2}$ = 50.563$\pm$0.025 days and its decay is dominantly proceeds through the emission of a $\beta$-particle with an endpoint energy of 1500.9 keV \cite{Sin13}.

       \subsection{The thin-window gas-cell target}

\begin{figure}[pos=htb]
                     \centering
                     \includegraphics[width=\linewidth]{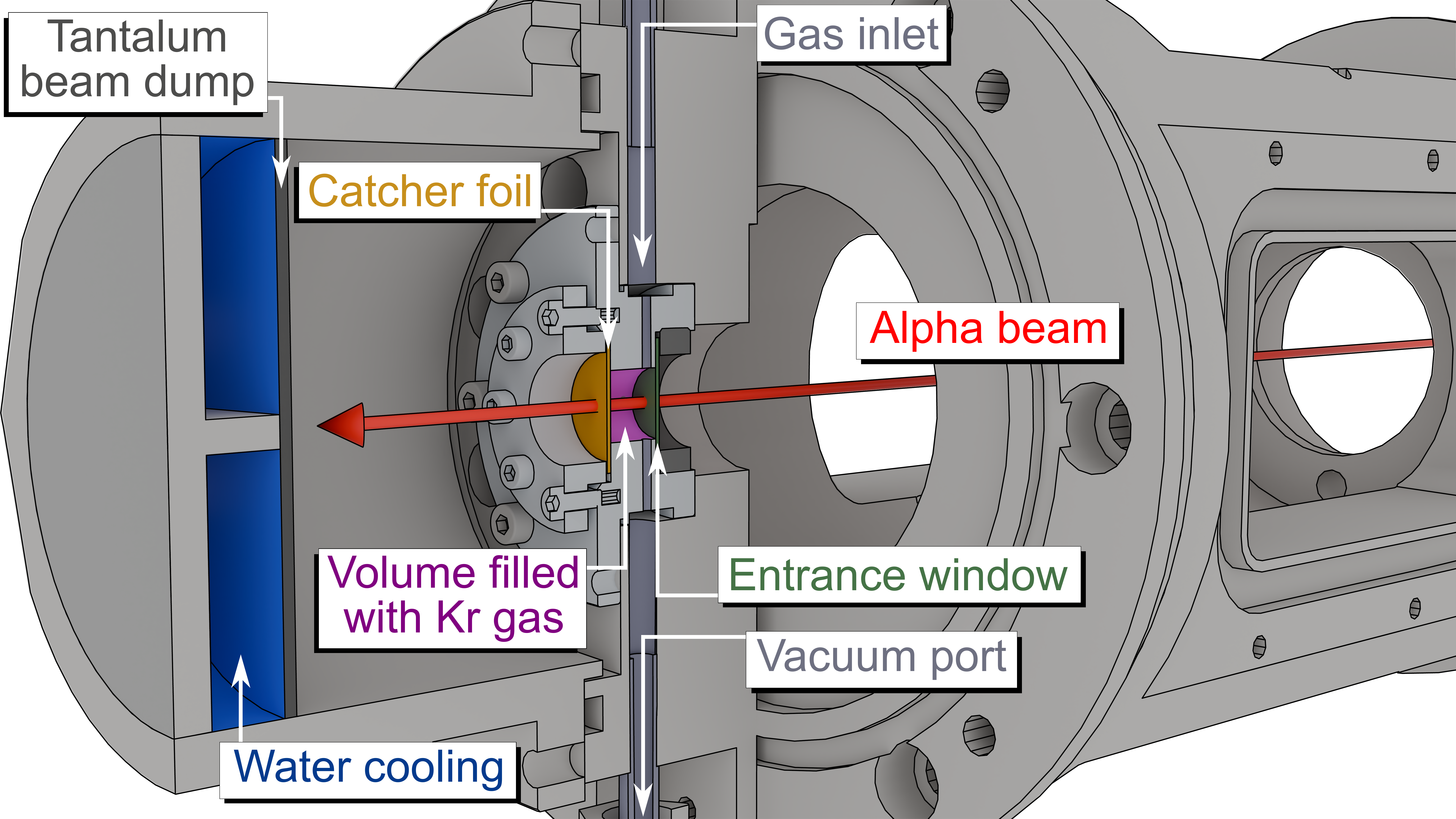}
                     \caption
                     {
                            Schematic view of the thin-window gas-cell target.
                     }
                     \label{fig:gasCell}
              \end{figure}

              A thin-window gas-cell target - shown schematically in Figure \ref{fig:gasCell} - was first used to measure the cross section. This chamber is a similar version to the one described in details in references \cite{Tot23}. 
              The thickness of the aluminum entrance foil, determined offline before the irradiation by measuring the energy loss of alpha particles emitted by $^{239}$Pu, $^{241}$Am and $^{244}$Cm isotopes, was found to be 10.29 $\pm$ 0.08 $\mu$m. 
              At the beginning of the irradiation, the cylindrical gas cell with a diameter of 12 mm and a length of 10.02 mm was filled with natural isotopic composition krypton gas at a typical pressure of 10 torr and thus the typical areal density of the $^{86}$Kr atoms was 5.7 $\times$ 10$^{16}$ atoms/cm$^2$. 
              The gas volume is sealed with a light-tight 10 $\mu$m thick high purity (99.999\%) aluminum exit foil purchased from Goodfellow GmbH. 
              Impurities in the foils have been studied by Inductively Coupled Plasma Mass Spectrometry (ICP-MS) technique. It was found that the foils contain a few ppm iron, cobalt, nickel and copper and the distribution of this contamination was found to be uniform within 2\% between each foil.
              The $^{89}$Sr isotopes produced by the $^{86}$Kr($\alpha$,n) reaction are implanted into this (so-called catcher) foil. 
              During the irradiation, the pressure slowly increases and thus the probability of successful implantation decreases. 
              This effect can be explained by the outgassing of the foils and the inner cell walls due to heat deposition. 
              The pressure in the gas target was monitored with a pressure gauge at 10-minute intervals and the implantation probability was calculated\footnotemark -- similarly to our previous works \cite{Tot23,Szu19} -- using the GEANT4 toolkit \cite{Ago03,All06,All16} for each period and found to be always larger than 80\%. 
              Furthermore, when the gas pressure reached about 20 torr, the irradiation was stopped, the gas target was emptied and then refilled with krypton gas at a pressure of approximately 10 torr. 
              For this reason, irradiations exceeding 12 hours had to be interrupted once or twice depending on the beam energy and current. 
              The energy loss in the entrance window and in the krypton gas was calculated using the SRIM code \cite{Zie10} and was found to be typically 0.92 MeV - 1.21 MeV (for the entrance window) and only a few keV (for the krypton gas), respectively.

\footnotetext{In these calculations not only the stopping of the reaction product in gas, but also implantation into the side walls of the chamber was investigated.}

              After the irradiation, the catcher foil was removed from the gas target and its activity was measured. 
              However, this $\beta$-activity measurement includes both the particles from the $\beta$-decay of the $^{89}$Sr reaction product and from any activity induced in the catcher foil. 
              Therefore, for each cross-section data point measurement, two irradiations were performed: with and without (called as background measurements) filling the thin-window gas cell target with krypton gas.

       \subsection{Production and properties of the implanted \texorpdfstring{$^{86}$Kr}{86Kr} targets}

\begin{figure}[pos=htb]
                     \centering
                     \includegraphics[width=0.9\linewidth]{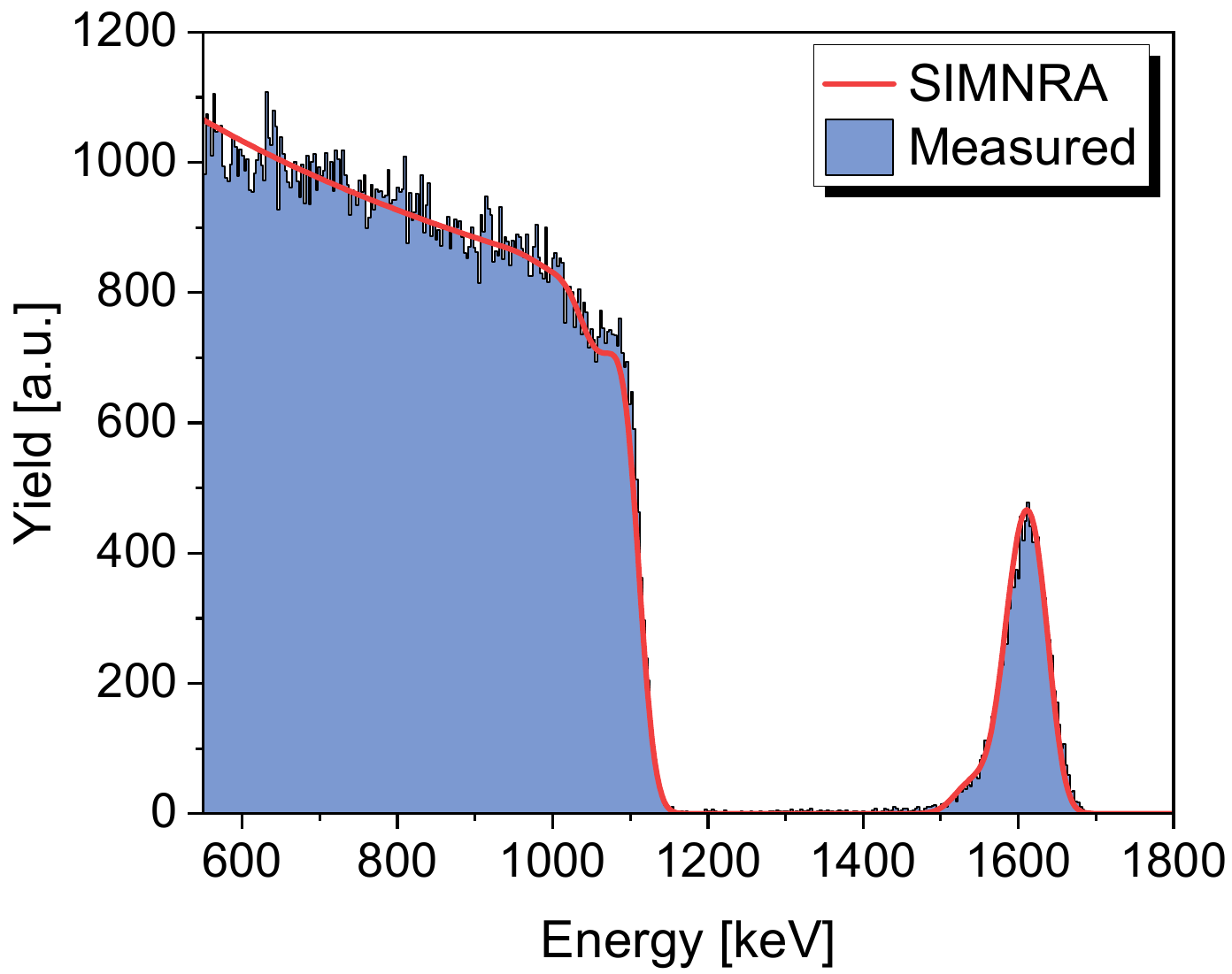}
                     \caption[]
                     {
                            Rutherford Backscattering Spectrum of an implanted target. 
                            The peak at 1600 keV is associated with the $^{86}$Kr content of the foil and aluminum edge is around 1100 keV.
                     }
                     \label{fig:RBS}
              \end{figure}

              Targets were also prepared by implanting $^{86}$Kr isotopes into a similar type of high purity aluminium foil. 
              The implantation was performed at the Heavy Ion Cascade Generator of the HUN-REN Wigner Research Centre for Physics, Budapest, Hungary. 
              The 200 keV krypton beam with a typical beam spot of 1 mm$^2$ was scanned vertically and horizontally over the entire sample surface in order to achieve good irradiation homogeneity within the exposed area.
              The current of the scanned ion beam was around 0.15 $\mu$A, which has been measured by a Faraday cup, and thus the nominal fluence of the implantation was about 5 × 10$^{16}$ atoms/cm$^2$. 
              In order to determine the implanted fluence, a silicon sample was also irradiated simultaneously with the Al foils. 
              The fluences were accurately determined using RBS on the silicon samples prior to the cross-section measurements. 
              The 2 MeV $^{4}$He$^+$-RBS measurement was performed in a scattering chamber equipped with a two-axis goniometer connected to the 5 MV Van de Graaff electrostatic accelerator of the HUN-REN Wigner Research Centre for Physics. 
              The beam was collimated to 0.5 × 0.5 mm$^2$ using two sets of four-sector slits. 
              The current was typically 10 nA, measured with a transmission Faraday cup \cite{Pas90}. 
              The typical dose was 4 $\mu$C. 
              Two tilt angles (7$^{\circ}$ and 60$^{\circ}$) were used to measure all samples, and RBS spectra were evaluated using the RBX code \cite{Kot94}, assuming the same layer structure. 
              Based on the RBS measurements, the Kr fluences were found to be between 4.92 × 10$^{16}$ Kr/cm$^2$ and 5.61 × 10$^{16}$ Kr/cm$^2$, derived with an accuracy of 2\% \cite{Jey12}. 
              After the reaction cross-section measurements performed at ATOMKI, the thicknesses of the targets were determined again by RBS to check the sample stability. For the RBS measurement a 2.0 MeV $\alpha$-beam provided by the Tandetron accelerator of ATOMKI was used. The beam enters the chamber through a 5 mm diameter collimator. The scattered $\alpha$-particles were detected by a collimated ion-implanted Si detector mounted in horizontal plane at a backward angle of 165$^{\circ}$ with respect to the beam direction. The typical beam currents were 150 nA (corresponding to 75 mC dose) and for calibration purposes aluminum, nickel and tantalum standards were used. A measured RBS spectrum together with that fitted with SIMNRA \cite{May97} is shown in Figure \ref{fig:RBS}. 
              The agreement between the two RBS measurements was found to be always better than 6\%, typically well within 3\% and the difference between the results of the two RBS measurements were considered as thickness uncertainty for each target.

\section{Irradiation and measurement of the induced activities}\label{sec:expSetup}

       To study the $^{86}$Kr($\alpha$,n)$^{89}$Sr reaction, $^{4}$He$^{++}$ beams with energies between 7.5 and 12.5 MeV (in steps of 1 MeV) were provided by cyclotron accelerator of ATOMKI.
       The same irradiation chamber was used as for the previous cross-section measurements (see e.g. \cite{Szu19,Kis21,Sze21,Tot23}). 
       The duration of the irradiations varied between 12 h and 36 h with beam currents of 0.4 - 0.9 $\mu$A. 
       The number of bombarding $\alpha$-particles was derived by current measurement, recording the integrated charge in every 1 min time interval.

       After irradiation, the foils containing the reaction products were removed from the irradiation chamber and transported to a detector setup for off-line decay counting. 
       The setup consisted of a 2 mm thick silicon particle detector with an active area of 50 mm$^2$ purchased from ORTEC (Model BA-017-050-2000 A-Series Partially Depleted Silicon Surface Barrier Detector). 
       This detector thickness is adequate to completely stop the electrons emitted by the widely used $^{90}$Sr-$^{90}$Y calibration source (the endpoint energies are 545.9 keV and 2278.5 keV \cite{Bas20}) as well as those emitted by the $^{89}$Sr reaction product (the endpoint energy is 1500.9 keV \cite{Sin13}). 
       The irradiated samples were placed in a plastic holder, at a distance of 4 mm from the surface of the particle detector. 
       The ($\alpha$,n) and ($\alpha,\gamma$) products of the reactions on other krypton isotopes and on the impurities of the catcher foil decay to negligible levels typically within a few days to a week. To be on the safe side, this minimum required time was roughly doubled and the $\beta$-particle countings were started 10 days - 2 weeks\footnotemark after the end of the irradiation.	
			
		\footnotetext{Due to their low natural abundances and / or the decay parameters of alpha-induced reaction products on them, other krypton isotopes were not disturbing for the activity measurement.}

\begin{figure}[pos=htb]
              \centering
              \includegraphics[width=0.9\linewidth]{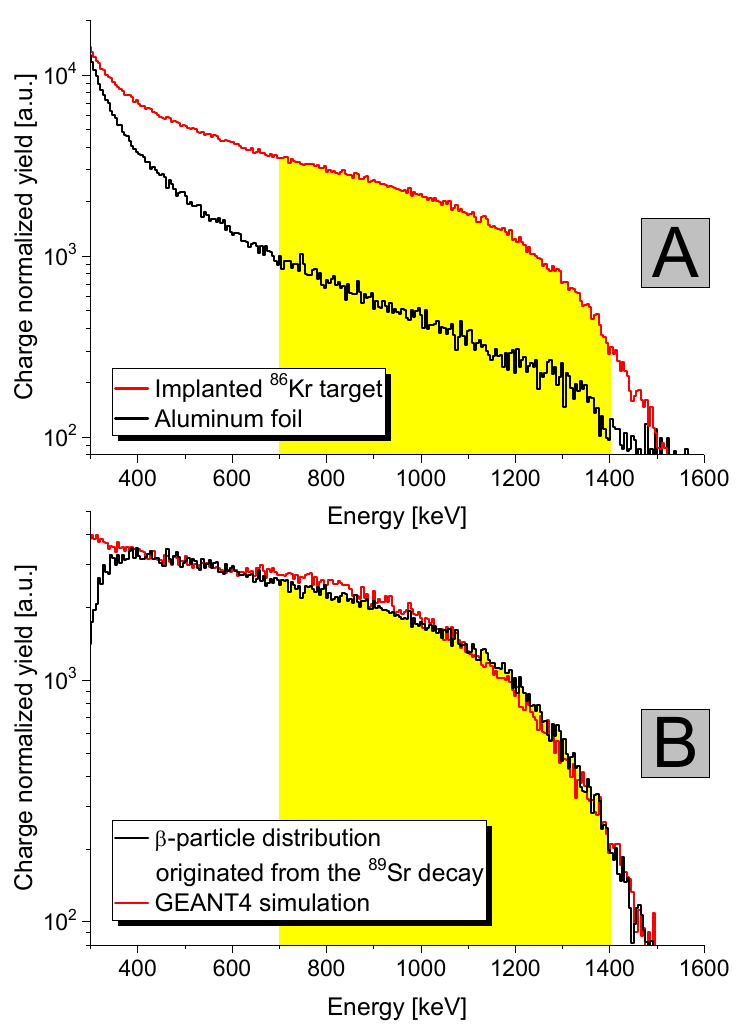}
              \caption[]
              {
                     Upper panel (A): a $\beta$-particle energy spectrum of a pure aluminum foil (black) and of an aluminum foil containing implanted $^{86}$Kr nuclei (red) after alpha irradiation with an energy of E$_{\alpha}$ = 8.5 MeV. 
                     Lower panel (B): the subtracted spectrum, corresponding to the distribution of electrons emitted by the $\beta$-decay of $^{89}$Sr reaction product. 
                     The yellow area indicates the region of interest.
              }
              \label{fig:spectra-2panel}
       \end{figure}

       Two irradiations were performed for each cross-section data point: one with krypton gas in the gas cell or with implanted targets and another one with an empty gas cell or a pure aluminum foil.
       This second irradiation was used to derive the background characterizing the measurement. 
       The irradiations and the $\beta$-particle countings were carried out under the same conditions in both cases and the $\beta$-spectra were normalized for the number of incident $\alpha$-particles. 
       The energy distribution of the electrons originating from the $\beta$-decay of the $^{89}$Sr reaction product was then determined by subtracting the background spectra. 
       The upper panel of Figure \ref{fig:spectra-2panel} shows the $\beta$-spectrum measured after irradiation with E$_{\alpha}$ = 8.5 MeV alpha beam on aluminum foils with and without containing implanted $^{86}$Kr atoms. 
       The lower figure shows the subtracted spectrum corresponding to the distribution of electrons from the $\beta$-decay of the $^{89}$Sr isotope which shows a good agreement with the GEANT4 simulated spectrum.
       
\begin{figure}[pos=htb]
              \centering
              \includegraphics[width=0.9\linewidth]{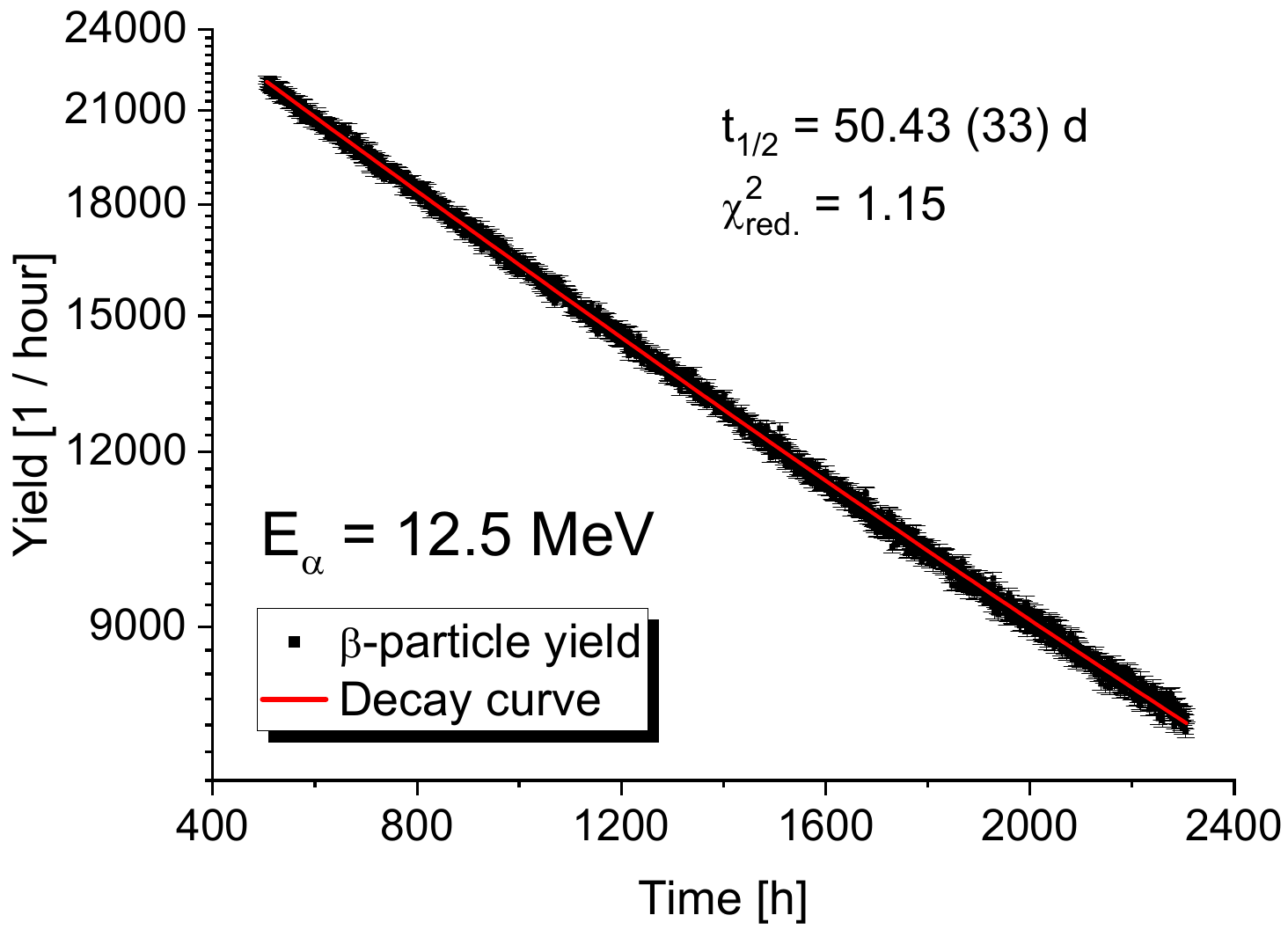}
              \caption[]
              {
                     The decay curve of $^{89}$Sr used to determine the cross section at E$_{\alpha}$ = 12.5 MeV.
              }
              \label{fig:half-1}
       \end{figure}

       A certain -- not negligible -- fraction of the low-energy $\gamma$-radiation, which e.g. originates from possible impurities, produces a signal in a 2 mm thick silicon detector. 
       To avoid this effect, $\beta$-particles only in the energy range E$_{\beta}$ = 700 keV - 1400 keV were considered during the analysis. 
       To verify this choice, for the E$_{\alpha}$ = 12.5 MeV irradiation the yield measurement was carried out for more than 3 months. 
       An exponential function was fitted to the $\beta$-particle yields determined as described above. 
       The half-life value determined in this way (t$_{1/2}$ = 50.43 $\pm$ 0.33 d) agrees well with the literature value (t$_{1/2}$ = 50.563 $\pm$ 0.025 d) \cite{Sin13}. 
       Figure \ref{fig:half-1} shows the decay curve together with the fitted exponential function of the $\beta$-particles emitted by $^{89}$Sr. 

              \subsection{Determining and verifying the detection efficiency}

\begin{figure}[pos=htb]
                            \centering
                            \includegraphics[width=0.9\linewidth]{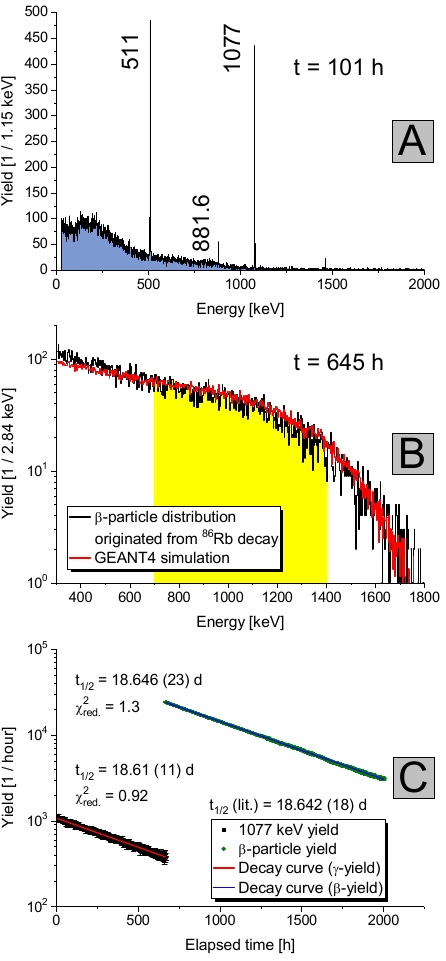}
                            \caption[]
                            {
                                   $\gamma$- (A) and $\beta$- (B) spectrum measured at t = 101 h and t = 645 h after irradiation of an aluminum foil containing $^{86}$Kr atoms with an E$_p$ = 5.3 MeV proton beam. 
                                   On panel A, the E$_{\gamma}$ = 1077 keV gamma peak -- originating from the decay of the $^{86}$Rb nucleus -- and the E$_{\gamma}$ = 881.6 keV peak originating from the $^{84}$Rb isotope, produced by the (p,n) reaction on the $^{84}$Kr contaminant are marked. 
                                   The lower figure shows the decay curves determined from $\gamma$-radiation and $\beta$- particle yield measurements. 
                                   The fitted exponential functions and the derived half-life values are also shown.
                            }
                            \label{fig:half-3}
                     \end{figure}

                     It has already been shown for several types of semiconductor particle detectors (HPGe, PIPS, SSB) that the GEANT4 toolkit can be used to simulate the detector response function even for low energy $\beta$-particles \cite{Sot13,H24}.
                     We performed similar simulations to determine the detection efficiency of our previously described $\beta$-counting setup using a calibrated $^{90}$Sr-$^{90}$Y source purchased from Eckert \& Ziegler Nuclitech GMBH.
                     The activity of the source at the time of the efficiency measurement was 1.103 $\pm$ 0.040 kBq.
                     Since the energy distribution of the $\beta$-particles emitted by the calibration source is well known, the number of particles with energies between E$_{\beta}$ = 700 keV and 1400 keV was calculated and compared with the measured spectra.
                     This energy range also serves as the region of interest, with good agreement between simulation and measurement, for the $\beta$-particle yield measurements of the $^{89}$Sr and $^{86}$Rb products, as shown in panel B of Figures \ref{fig:spectra-2panel} and \ref{fig:half-3}.

                     The detector efficiency determined above was verified as follows. 
                     Using an implanted $^{86}$Kr target, a strong $^{86}$Rb source was made by the $^{86}$Kr(p,n)$^{86}$Rb reaction. 
                     This rubidium isotope has a half-life of 18.642 $\pm$ 0.018 days and decays by $\beta^-$-decay with Q$_{\beta}$ = 1776 keV. The $\beta$-decay of the nucleus leads either to the first excited state of the daughter nucleus or to the ground state.
                     In the first case the decay is accompanied by the emission of a $\gamma$-ray with an energy of E$_{\gamma}$ = 1077.0 keV \cite{Neg15}. 
                     The well-known relative intensity of this $\gamma$ radiation (I$_{\gamma}$ = 8.64 $\pm$ 0.05\%) allows the determination of the source activity. 
                     The $\gamma$-emission was followed with a HPGe detector for 670 hours, the spectra were saved every hour and the measured yields were fitted with an exponential function.
                     The derived half-life (t$_{1/2}$ = 18.61 $\pm$ 0.11 d) is in good agreement with the more precise literature value. 
                     Knowing the peak areas, the characteristics of each $\gamma$-counting (start time and length), the decay parameters (t$_{1/2}$, I$_{\gamma}$) and the detector efficiency, the activity of the source was derived (and found to be 0.652 kBq at the start of the activity measurement). 

                     The $\beta$-decay of the $^{86}$Rb nucleus occurs via the emission of $\beta$-particles with end point energies of either E = 699.2 keV (I$_{\beta}$ = 8.64 $\pm$ 0.04 \%) or E = 1776.2 keV (I$_{\beta}$ = 91.36 $\pm$ 0.04\%). 
                     After the end of the $\gamma$ counting the source was placed in front of the silicon particle detector, where the yield of the $\beta$-particles were measured for more than 55 days and the spectra were saved every hour. 
                     A pure aluminum foil was also irradiated and its $\beta$-spectrum measured as described above.
                     By subtracting this background spectrum, the spectrum of the $\beta$-particles originating from the $^{86}$Rb isotope was determined for each hour-long interval. 
                     The middle panel of Figure \ref{fig:half-3} shows the energy distribution of $\beta$-particles originated from the decay of $^{86}$Rb measured at t = 645 h after the start of the $\gamma$-count. 
                     For the analysis, a fraction of the $\beta$-particles belonging to the ground state decay were used by selecting the events with energies between E$_{\beta}$ = 700 keV and 1400 keV. 
                     The yields of the selected $\beta$-particles, measured in one-hour intervals, were determined by numerical integration and fitted with an exponential function. 
                     The derived half-life (t$_{1/2}$ = 18.646 $\pm$ 0.023 d) is in perfect agreement with the literature value, thus confirming that the measured particles belong to the decay of $^{86}$Rb. 
                     The lower panel of Figure \ref{fig:half-3} shows the decay curves and the exponential fits. 
                     Knowing the number of $\beta$-particles, the I$_{\beta}$ relative intensity, the timing of the activity measurement, and the activity of the source determined by $\gamma$-counting, the efficiency characterizing the $\beta$-yield measurement was determined. 
										As a result of the above discussed procedures, the detection efficiency was found to be 14.45 $\pm$ 1.16\%.

                     The proton irradiation revealed that the targets made by implanting $^{86}$Kr isotopes contain a small but not negligible amount of $^{84}$Kr isotopes, as seen from the 881.6 keV $\gamma$-peak in panel A of Figure \ref{fig:half-3}. 
                     The amount of this contaminant was determined on the basis of the known efficiency of the HPGe detector, the decay parameters, the characteristics of the activity measurement, and the cross section calculated with the NON SMOKER model (as a conservative estimate, we assigned an uncertainty of 50\% to this quantity). 
                     As a result, we found that our samples contain a maximum of 4\% $^{84}$Kr isotope. 
                     And this possible impurity content was taken into account as a systematic error in the uncertainty of the target thickness determined by the RBS technique.

\section{Studying the \texorpdfstring{$^{86}$Kr($\alpha$,n)$^{89}$Sr}{86Kr(alpha,n)89Sr} reaction}\label{sec:86Kr}

To test the performance of the above detailed setup and experimental approach, two irradiation were carried out to measure the cross section
of the $^{86}$Kr($\alpha$,n)$^{89}$Sr reaction. The first irradiation was carried out using the thin-window gas cell target, the beam energy was E$_{\alpha}$ = 12.5 MeV. For the 2nd irradiation an implanted target was used and the beam energy was chosen so that the center-of-mass energy was similar to the previous irradiation (E$_{c.m.}$ = 11.07 $\pm$ 0.05 MeV and 11.06 $\pm$ 0.05 MeV, respectively). The resulted cross sections were found to be 88.40 $\pm$ 12.53 mbarn (measured using the thin-window gas cell target) and 86.35 $\pm$ 12.27 mbarn (measured with implanted target). The uncertainty of the cross sections is the quadratic sum of the following partial errors: efficiency of the $\beta$-particle detection (8\%), determination of the target thickness (5-8\%), current measurement (3\%), and counting statistics (8-12\%) (the uncertainty of the decay parameters is negligible). The effective center-of-mass energy takes into account the energy loss of the beam in the target and in the entrance foil, its quoted uncertainty corresponds to the energy calibration of the $\alpha$-beam and to the uncertainty of the energy loss calculated with the SRIM code \cite{Zie10}. The two cross sections measured at the same effective energy with different targets are within few percent.

\begin{figure}[pos=htb]
                            \centering
                            \includegraphics[width=0.9\linewidth]{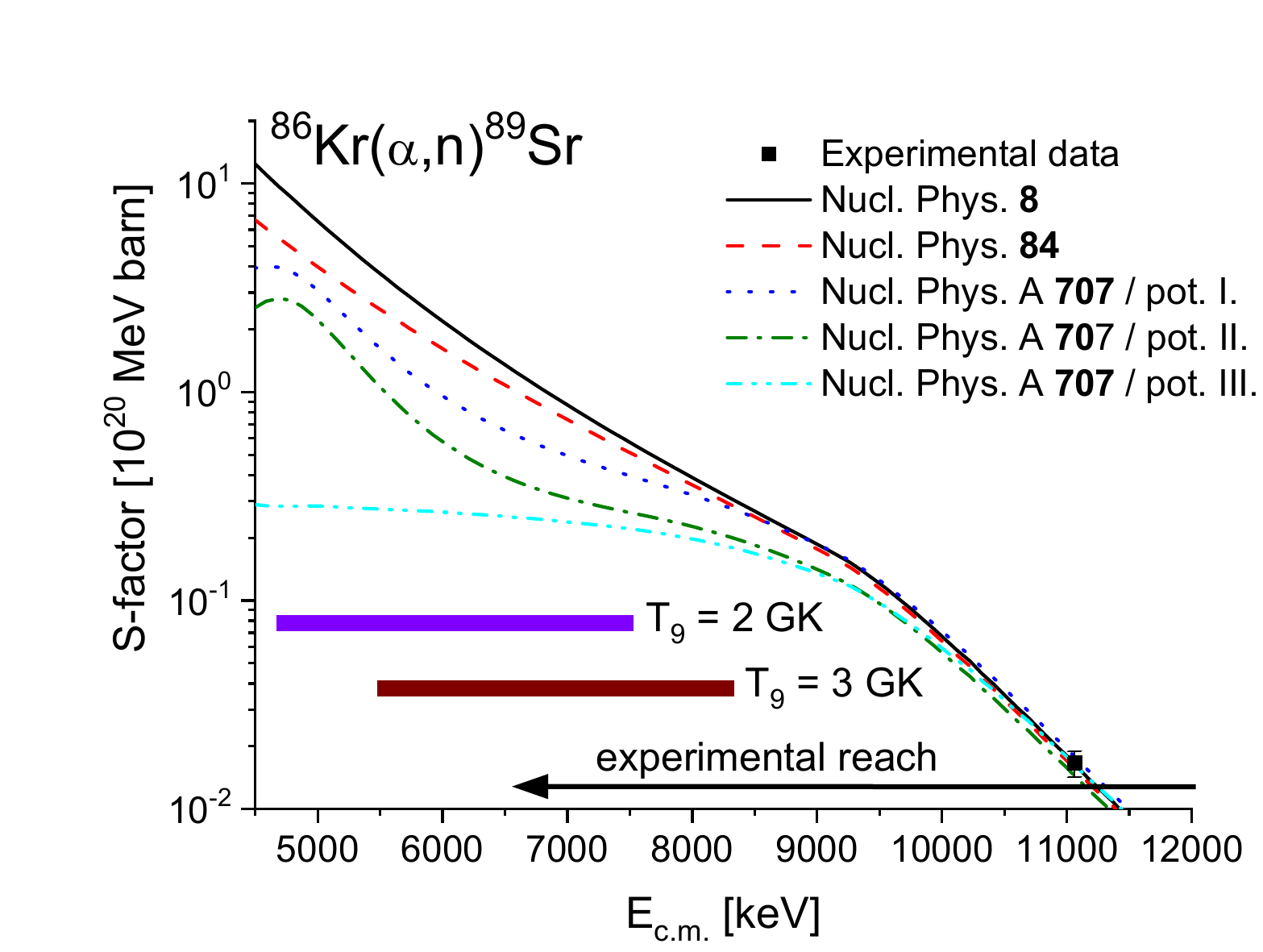}
                            \caption[]
                            {Predictions, calculated using different global $\alpha$-nucleus optical model potentials \cite{Wat58,McF66,Dem02}, for the $^{86}$Kr($\alpha$,n)$^{89}$Sr reaction S-factors compared to the new experimental data.}
                            \label{fig:s-factor}
                     \end{figure}

The astrophysical S-factors\footnotemark of the $^{86}$Kr($\alpha$,n)$^{89}$Sr reaction are compared to theoretical calculations using various $\alpha$-nucleus optical potentials \cite{Wat58,McF66,Dem02} implemented with the TALYS code \cite{Kon23} on Figure \ref{fig:s-factor}. While at higher energies the theoretical cross section values agree with each other and also with the experimental results, in the astrophysically relevant energy region, the theoretical predictions differ significantly, exceeding at the lowest energies even an order of magnitude. Thus, more data is clearly needed to draw reliable conclusions and for this purpose, series of measurements are underway using the setup described in this manuscript.. The results of these measurements as well as the theoretical and astrophysical analysis of the results are planned to be published elsewhere.
 
\footnotetext{In astrophysical application instead of the highly energy dependent cross section $\sigma(E)$ often the astrophysical S-factor S(E) is used which has a smoother energy dependence. The S(E) is defined as: $S(E)\equiv\frac{E}{\exp{(-2 \pi \eta)}}\sigma(E)$ where $\eta$ is being the Sommerfeld parameter. For more details, see \cite{ilibook}.}

\section{Summary and outlook}

       Alpha-induced reactions play a particularly important role in the astrophysical weak r-process, which synthesizes the neutron-rich isotopes between strontium and silver. In the present work, our experimental method used to measure the yield of the electrons emitted during the $\beta$-decay of the $^{89}$Sr activation product was introduced. For the presented measurements, we used both a thin window gas cell target and targets made by implanting $^{86}$Kr atoms into aluminum foils. A 2 mm thick silicon ion implanted detector was used to measure the $\beta$-particle yields, the efficiency of which were determined using a $^{90}$Sr source, GEANT4 simulations and verified using the $^{86}$Kr(p,n)$^{86}$Rb reaction. 
       
	Currently, the study of the $^{86}$Kr($\alpha$,n)$^{89}$Sr reaction was completed at ATOMKI using the setup and procedure described in this manuscript, and the preparations of the $^{87}$Rb($\alpha$,n)$^{90}$Y reaction cross section measurement have begun. Based on the irradiation and activity measurement parameters, cross-sections data will also be available in part of the astrophysically relevant energy region, thus the widely used optical potential parameter sets will be further constrained.

\section*{Acknowledgments}
       This work was supported by NKFIH (K147010, K134197, FK134845) and by the European Union (ChETEC-INFRA, Project No. 101008324).

\bibliography{Kr86-technicalArticle}

\end{document}